\documentclass{epl}

\usepackage{graphicx}
\usepackage{amsmath,amssymb}

\title{Block to granular-like transition in dense bubble flows}
\shorttitle{Block to granular-like transition}
\author{N.Vandewalle\inst{1} \and S.Trabelsi\inst{1} \and H.Caps\inst{1}}
\shortauthor{N. Vandewalle {\it et al.}}
\institute{
  \inst{1} GRASP, Dept. of Physics B5, University of Li\`ege, B-4000 Li\`ege, Belgium.
  }
\date{Received: \today / Revised version: date}


\begin{document}

\maketitle

\begin{abstract}
We have experimentally investigated 2-dimensional dense bubble flows underneath inclined 
planes. Velocity profiles and velocity fluctuations have been 
measured. A broad second-order phase transition between two dynamical regimes is 
observed as a function of the tilt angle $\theta$. For low $\theta$
values, a block motion is observed. For high $\theta$ values, the 
velocity profile becomes curved and a shear velocity gradient appears 
in the flow. 
\end{abstract}

\section{Introduction}

\begin{figure}[h]
\begin{center}
\includegraphics[width=8cm]{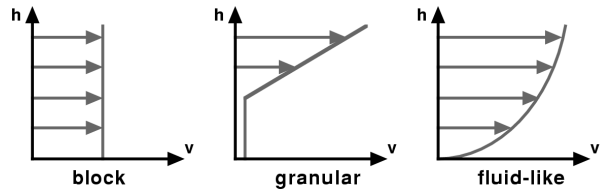}
\caption{Typical velocity profiles $v_{h}$ encountered in dense flows. From left to right: block motion, surface 
flow and Poiseuille flow.}\label{profiles}
\end{center}
\end{figure}

The dense granular flow of identical particles is a complex phenomenon which 
involves multiple processes such as friction, correlated motion and 
inelastic collisions \cite{duranbook,pouliquen,prlpoiseuille}. Various flow regimes 
have been identified in different experiments. Three major flow 
regimes are emphasized in Figure~\ref{profiles}. Block motion occurs for example 
when particles form some jammed phases \cite{jam1,jam2,bonamy_gm}. In granular 
flows on a pile, only the surface grains are in motion. This rolling 
phase has typically a thickness of 10-20 grains and the velocity $v_h$ as a 
function of the height $h$ is roughly linear for 
that surface grains \cite{rajchenbach,azanza,scaling}. 

A lot of efforts has been devoted to the understanding of dense flows 
of particles. Some studies \cite{contact,grest1} address the problem of 
contact force networks, i.e. the presence of arches, which plays an 
important role \cite{role}. Fluctuations and correlations of the 
grain motion as a function of the shear rate have been also studied. 
Collective behaviors have been put into evidence such as the motion 
of clusters of grains \cite{bonamy_gm,bonamy}. Different theoretical approaches have been developed for describing 
dense flows. Among them, one can cite the Saint-Venant approach which 
considers some conservation equations in the rolling phase 
\cite{saintvenant}. Pouliquen and coworkers \cite{pouliquen} have 
also proposed that the granular flows can be considered as 
quasi-static flows, the global motion resulting from fractures. 
Nevertheless, no general theory can predict the occurrence of the different flow regimes and corresponding 
velocity profiles from the various physical parameters such as the 
density, the shear rate and the friction.

In 1947, Bragg and coworkers proposed a simple experiment 
\cite{bragg} in order to put into evidence some crystallographic 
ordering. This experiment is also famous because it was quoted in the 
Feynman's lectures \cite{feynman}. The experiment consisted in 
producing small identical bubbles at the surface of a liquid. 
Hexagonal packed structures appeared and crystallographic domains 
separated by grain boundaries were observed. Recent experiments \cite{tamtam,dimeglio} 
were devoted to the study of defects in such bubble rafts. 

We propose to modify the Bragg's experiment in order to study dense 
flows of identical bubbles. The flow of bubbles is of physical 
interest since contacting bubbles are characterized by a nearly zero friction \cite{nicolson} ! 
Our experimental system corresponds thus to an ``ideal'' 
dense flow. We show in this letter that this system of bubbles 
presents a phase transition between two distinct dense flow regimes 
depending on the kinetic energy of the incoming bubbles.

\section{Experimental setup}

\begin{figure}[h]
\begin{center}
\includegraphics[width=6cm]{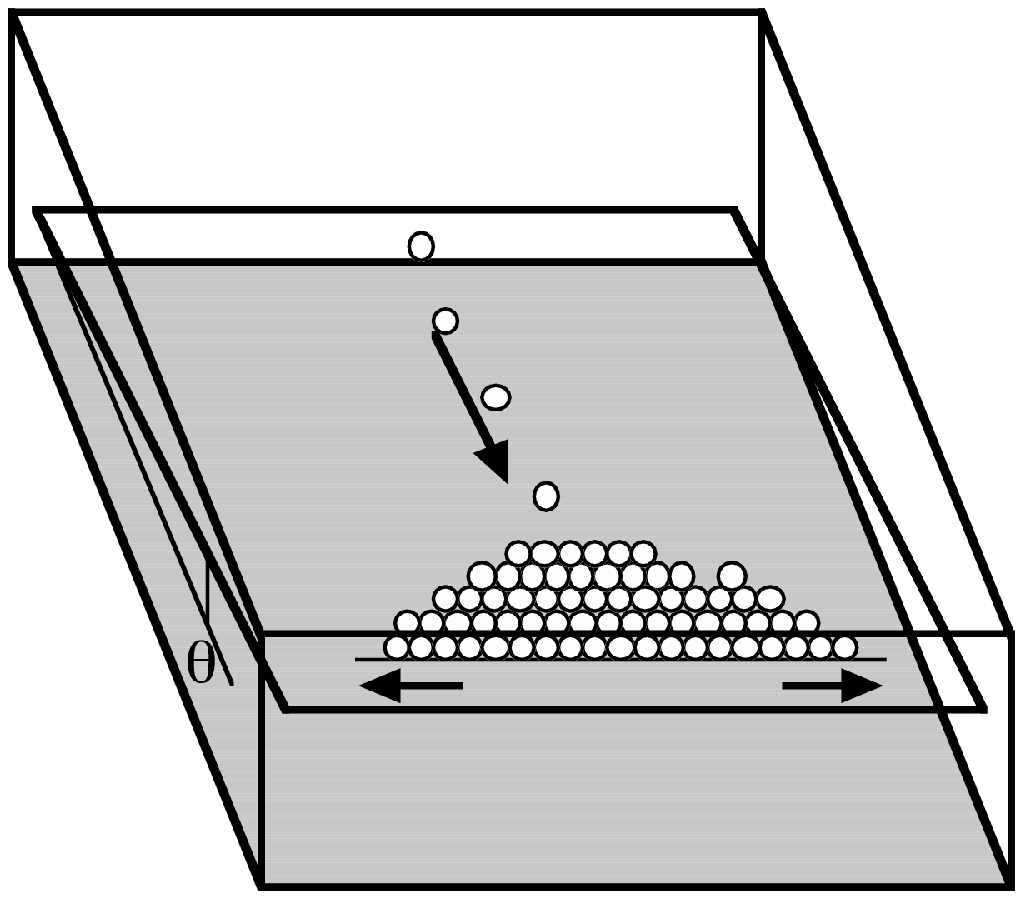}\qquad\qquad\includegraphics[width=6cm]{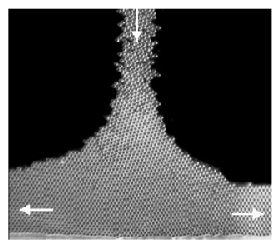}
\caption{A transparent inclined plane is immersed into water and tilted by an angle $\theta$. Small air 
bubbles are injected from below at the bottom of the plane and rise towards the 
top where they aggregate on a linear rough obstacle. A sketch of the experimental setup (left) 
and typical bubble packing taken by the CCD camera placed above the setup (right) are illustrated. 
The angle is $\theta = 1^\circ$. Arrows indicate the motion of bubbles.}\label{setup}
\end{center}
\end{figure}

Our experiment is analogous to the filling of a 2D-silo by
continuous feeding of granular material from a central hopper. The
experimental setup consists in a transparent inclined plane which is 
immersed into water [see Fig.~\ref{setup}]. The tilt angle $\theta$ is small and ranges from $0^\circ$ 
to $3^\circ$ in our experiments. The error on $\theta$ is $0.1^\circ$. 
Spherical air bubbles are injected from below at the bottom of the tilted plane. The bubble production 
rate is kept constant at $5$ bubbles by one second. The size of the bubbles can be controlled by 
the air pump. The bubble radius is typically $R=0.75$ mm in the present 
study. In order to avoid the coalescence of the bubbles, a small 
quantity of commercial soap (based on SDS surfactants \cite{broze}) is added to the water. 
The mixture has a surface tension $\gamma=0.03$ N/m 
and a viscosity $\eta=0.001$ kg$\,$m${}^{-1}\,$s${}^{-1}$. Due to buoyancy 
$B$, the bubbles rise beneath the inclined plane. A rough rigid object 
has been placed at the top of the plane such that bubbles aggregate 
there. The obstacle roughness is typically one bubble diameter and 
avoids purely translational motions of the bubble stack
\cite{grest3}. One should note that in the direction perpendicular to
the plane, the bubble stack is only one layer of bubbles thick. We
thus have a really 2D bubble pile. A CCD camera 
captures top views of the packing of bubbles through the transparent
tilted plane. 

The unique parameter to be investigated is the tilt angle $\theta$ of the plane. 
Indeed, the angle $\theta$ controls the 
buoyancy
\begin{equation}
B = {4 \pi R^3\Delta \rho g \over 3}\sin \theta,
\end{equation}
where $\Delta \rho \approx 1000$ kg\, m$^{-3}$ is the difference 
between air and liquid densities. One should also note that the 
bubble motion is mainly limited by the viscous drag force
\begin{equation}
  f \sim \eta R v_i.
\end{equation} 
The bubble/plane friction is small with respect to the drag 
force $f$ as we will see below. Force equilibrium $B=f$ gives the 
terminal incoming velocity $v_i$ of the bubbles which is a few millimeters per 
second since $\sin \theta \approx 10^{-2}$ in our experiments. We are
thus in a laminar regime ($Re\approx 1$).

\section{Experimental results} 

Bubbles tend to aggregate in an ordered hexagonal packing since the 
bubble size is monodisperse. The packing is continuously fed by 
new bubbles. As a consequence, the packing evolves and a dense flow 
of bubbles is seen in both tails. As for usual granular flow, a thin
layer of moving grains is observed \cite{nasuno}. Moreover, smaller
motions are also observed deeper inside the packing, even for low $\theta$ values. 

Figure~\ref{setup} presents a typical packing of bubbles for $\theta=1^\circ$. 
The shape of the pile is clearly rounded. If the incoming bubble flow is 
interrupted, the bubbles slide on each other under the
influence of buoyancy and the angle of the bubble stack tends to
zero. The pile collapses completely. This indicates that 
the friction between contacting bubbles is quite low, as discussed in the introduction. 
As contacting bubbles are sliding on each other, small shape deformations 
are also observed at the bottom of the pile. Thus, the bubble pile is not a strict granular 
medium in the sense that no arch is created.

\begin{figure}[h]
\begin{center}
\includegraphics[width=6cm]{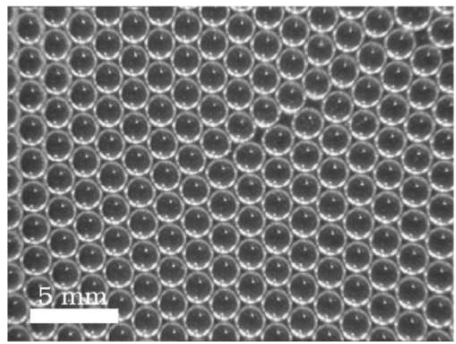}
\caption{Part of the packing. Bubbles tend to 
aggregate in a hexagonal packing. A dislocation is seen in the center 
of the picture.}\label{pack}
\end{center}
\end{figure}

In the packing, the bubble motion is mainly due to moving dislocations [see Fig.~\ref{pack}]. 
Those punctual defects are created at the surface of the packing and 
propagate along bubble lines at various speeds. 
One should note that the mean dislocation speed is larger than the bubble speed. This 
is due to the cooperative motion of the bubbles. For larger values of 
$\theta$, several dislocations are seen in the different parts of the pile. 
The majority of the bubbles is in motion. 

\begin{figure}[h]
\begin{center}
\includegraphics[width=6.5cm]{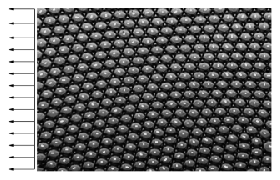}\qquad\includegraphics[width=6.5cm]{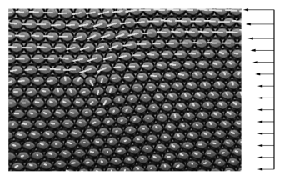}
\caption{Instantaneous velocity fields of a part of the tail of the bubble pile for (left) $\theta=0.5^\circ < \theta_c$ and (right) $\theta=1.5^\circ > \theta_c$. The bubbles are observed to move from right to left. A block motion is observed for $\theta< \theta_c$, while a velocity profile decreasing with the depth $h$ is found for $\theta> \theta_c$. Sketches of the velocity profiles are also illustrated.}\label{champ}
\end{center}
\end{figure}

In order to quantify and to measure the bubble flow, we have tracked the motion
of every bubble in the tail of the bubble pile. Movies of the evolution 
of different parts of the bubble stack have been recorded at a frame
rate of $12$ fps. The recorded
areas contains typically $30\times 30$ bubbles. A precision of $0.03\,$mm on bubble 
positions is obtained. 
We have then measured the bubble paths and deduced velocity profiles using a particle 
tracking algorithm. 
Figure~\ref{champ} presents two typical instantaneous ($\Delta
t=0.083$\, s) velocity fields in the tail of the bubble pile, 
for respectively $\theta=0.5^\circ$ and $\theta=1.5^\circ$. One can observe slow bubble motions for 
a small $\theta$ value (left) and moving layers of bubbles over a 
quasi-static phase for a larger $\theta$ value (right). Velocity measurements presented hereafter result 
from averages over long movies (more than $30\,$s). 


\begin{figure}[ht]
\begin{center}
\includegraphics[width=6cm,height=5cm]{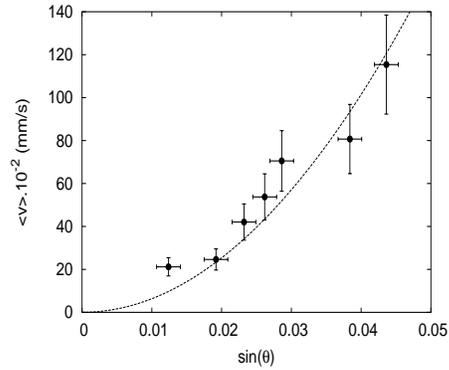}
\caption{Mean velocity $\langle v\rangle$ as a function of $\sin(\theta)$. Each dot 
represents an average over $1500$ pictures and over $1.3\ 10^6$ bubbles. Error bars are indicated. The 
curve is a fit emphasizing a $\sin^2(\theta)$ behavior.}\label{activ}
\end{center}
\end{figure}

Figure~\ref{activ} presents the {\it mean velocity} $\langle v\rangle$ in the 
bubble pile as a function of $\sin(\theta)$. We propose a quadratic
behavior of the mean velocity as emphasized by the fit in
Figure~\ref{activ}. The mean velocity $\langle v\rangle$ would then be controled by the
tilt angle $\theta$ and would be proportional to the kinetic energy ($\sim v^2_i$) of 
individual bubbles before they reach the packing.
\begin{figure}[hb]
\begin{center}
\includegraphics[height=5cm,width=6cm]{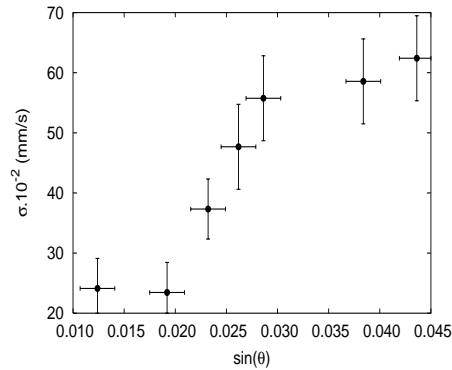}
\caption{Velocity fluctuations $\sigma$ defined by Eq.(\ref{eqT}) as a function of $\sin(\theta)$. A broad transition is seen at 
$\sin(\theta_{c})\approx 0.025$ or $\theta_c \approx 1.4^\circ$. Each dot represents an average over $1500$ pictures 
and over $1.3\ 10^6$ bubbles. Error bars are indicated.}\label{T}
\end{center}
\end{figure}
In order to characterize the bubble flow, we have also performed 
measurements of the velocity fluctuations, similarly to the granular
`temperature' \cite{azanza}. The velocity fluctuations amplitude $\sigma$ is then defined by
\begin{equation}\label{eqT}
\sigma = \sqrt{ \langle v^2 \rangle-  \langle v \rangle^2 } \, ,
\end{equation} 
i.e. the second moment of the velocity distribution in both $x$ and
$y$ directions \cite{azanza}. 
Those velocity fluctuations $\sigma$ are plotted as 
a function of $\sin(\theta)$ in Figure~\ref{T}. One can observe a `jump' 
of the velocity fluctuations $\sigma$ for tilt angles above $\sin(\theta_c)\approx 
0.025$. This suggests the presence of a broad second-order phase transition as a function 
of the tilt angle $\theta$. The critical angle corresponds to $\theta_c\approx 1.4^\circ$. Around this critical angle, two dense flow regimes should 
be distinguished. They will be characterized below.

For low angle values ($\theta < \theta_c$), the dislocations are 
created at the surface of the pile. They propagate in 
diagonal directions and are reflected at the bottom of the bubble 
packing. As a consequence, we have observed the motion of giant 
triangular domains of bubbles. Above the critical angle $\theta_c$, 
we observe numerous dislocations. The density of dislocations is 
so high that they are interacting by pairs. Indeed, we have observed the 
collision, the merging, and the annihilation of dislocation 
pairs. The cooperative motion of the dislocations lead to the motion 
of small clusters of bubbles.

\begin{figure}[h]
\begin{center}
  \includegraphics[width=7cm]{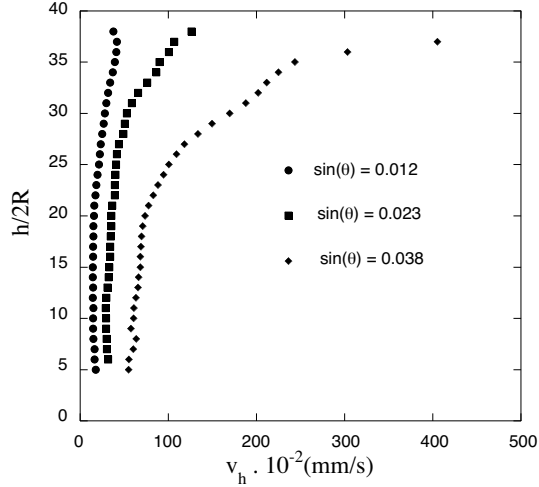}
\caption{From left to right: velocity profiles $v_{h}$ for respectively  $\sin(\theta)=0.012$, $\sin(\theta)=0.023$, and 
$\sin(\theta)=0.038$. Below $\theta_{c}$, we observe nearly constant vertical profile as 
in block motions, while above, we obtain a granular-like profile.}\label{prof}
\end{center}
\end{figure}

The above behaviors imply distinct velocity profiles. Velocity profiles 
have been measured as a function of the vertical position $h$ reduced 
by the bubble diameter $2R$ and counted from the obstacle. 
Three typical velocity profiles are given in Figure~\ref{prof}. Velocity profiles $v_{h}$ 
reveal two components: (i) a mainly constant profile which dominates 
for low $\theta$ values and (ii) a curved profile (a gradient) seen for 
high $\theta$ values. The curvature above $\theta_{c}$ is similar to 
the velocity profile of granular systems [see Fig~\ref{profiles}] :
a velocity profile decreasing abruptly in the more static phase (e.g. deeper in 
the pile) and growing in the flowing layer \cite{bonamy_gm}. The velocity profile penetrates 
 as deeply in the pile as the tilt angle $\theta$ value is large, as in granular systems \cite{bonamy_gm}.

The `vertical' profiles at low $\theta$ values can be explained in terms 
of block motions of the giant triangular domains of bubbles. Domains are sliced by 
fast moving dislocations. As the kinetic energy of the incoming 
bubbles increases, the size of the bubble domains decreases. 
The density of dislocations increases and the bubble stack becomes more `fluid'. At this stage, 
interacting dislocations which are created at the surface of the flow are observed to 
annihilate deep in the dense flow. The study of the interaction/annihilation between dislocations 
could explain the particular curvature of our velocity profiles. This needs more theoretical work and 
is outside the scope of this letter.

The `jump' in the velocity fluctuations $\sigma$ suggests that we have a transition between block and 
granular flows depending on the density of dislocations in the packing. 
Varying the bubble size $R$ or the bubble production rate does not affect 
qualitatively the results reported herein. Moreover, the bubble motions and 
dislocation mechanisms reported hereabove are also observed when changing the 
soap-water mixture.

In the future, the analogy between granular and bubble flows could
be developed by changing the geometry of the bubble flow and
considering ``chute flow''. By tilting the rough obstacle relatively
to the incomming bubble flow, one would get a ``chute'' flow of bubble
on an inclined substrate. Tuning the bubble rate production and the
tilt angle, could control the thickness of the flowing layer.   

\section{Summary}

We have studied bubbles moving 
and aggregating on inclined planes. We have underlined the similarities 
between dense bubble flows and granular flows. Dense bubble flows 
can be considered as `ideal' granular flows with nearly zero friction. We have emphasized a 
broad second-order transition between two dense flow regimes as a function of the 
tilt angle $\theta$ of the plane below which the bubbles are injected.

\begin{acknowledgements}
This work is financially supported by the contract ARC number 02/07-293.
HC thanks FRIA (Brussels, Belgium) for financial support. NV thanks 
A.Pekalski, O.Pouliquen, and S. Dorbolo for fruitful discussions. The authors are also 
grateful to G. Broze for surfactant informations (Colgate-Palmolive Research facility).
\end{acknowledgements}

\end{document}